\documentclass[a4paper]{article}
\usepackage{CJK}
\usepackage{amssymb}
\usepackage{amsthm}
\usepackage{graphicx}
\usepackage{booktabs}
\usepackage[boxed]{algorithm2e}
\usepackage{subfigure}
\usepackage{multirow}

\title{Privacy-Preserving Vehicular Announcements Aggregation Scheme Based on Threshold Ring Signature}
\author{Yichen Jiang\thanks{Y. Jiang is with the Department of Computer and Information Technology, Beijing Jiaotong University, Beijing, China, e-mail: (12112066@bjtu.edu.cn).}}

\begin{document}

\maketitle

\begin{abstract}
As the most promising application in VANETs, the vehicular announcement allows vehicles to send announcement messages about road conditions to other vehicles far away. The security requirements of reliability and privacy in the vehicular announcement are not easily achieved simultaneously due to the notorious sybil attack. In this paper, we present a novel privacy-preserving vehicular announcements aggregation scheme. The proposed scheme provides threshold authentication and flexible anonymity using message aggregation and interactive threshold ring signature which allows nondeterministic different signers to generate signature commonly in the environment not fully trusted. Different from existing works in an attack-then-trace mode, our scheme defends against the sybil attack beforehand. To our best knowledge, it is the first privacy-preserving scheme preventing the malicious users from launching the sybil attack in advance. Through extensive evaluation, we show the availability and the efficiency of the proposed scheme.
\end{abstract}


\section{Introduction}
The technology of Vehicular Ad Hoc Networks (VANETs) has drawn wide attentions from both academia and industry in recent years. VANETs are considered to be the most potential technology solving the problem of traffic safety and efficiency \cite{HL08}. In VANETs framework, vehicles are equipped with On Broad Unit (OBU) so as to communicate with each other (V2V communication), as well to roadside infrastructures (V2I communication) equipped with Roadside Unit (RSU).

The vehicular announcement, one of the most promising applications in VANETs, allows vehicles to sense the road condition and send the information (e.g. traffic jams, car accidents, road constructions) to other vehicles so that the receivers of the announcements can avoid the troublesome points in advance. The announcement messages will be disseminated in a wide range, so the information reliability is of great importance in this application. Majority-based authentication, or threshold-based authentication, is a common method to achieve reliability. The number of vehicles sending the same message will help the receiver believe the truthfulness of a certain announcement message. Message aggregation is an efficient way to implement majority-based authentication, which helps to reduce the forwarding of duplicated messages and the waiting time of the message receiver \cite{RAH06}.

Privacy issues in VANETs are widely concerned in recent years, and it is considered to be necessary to provide privacy-preserving environment in VANETs \cite{HL08}\cite{PP05}. The importance of privacy in VANETs is comprehensively analyzed in \cite{Dot06}. However, due to the notorious sybil attack\cite{Dou02}, the majority-base authentication is not properly functioning in privacy-preserving environment. As analyzed in many literatures \cite{CNW11}\cite{KWL09}\cite{Lin13}, naively combining anonymous techniques and majority-based authentication cannot satisfy the security requirements in privacy-preserving VANETs environment. It is still a challenging problem to implement the tradeoff between reliability and privacy. For instance, consider the following scenario:

Suppose Alice is driving on the main road, but the speed is extremely slow. She wants to issue a message telling the other drivers that there is a traffic jam on the main road, so the receivers can avoid the troublesome point. Only one person issuing the message cannot convince the other people. She needs the help of other drivers suffering from the same traffic jam to issue the message together. However, all these messages are sent anonymously in privacy-preserving VANETs. The receivers of the messages cannot be sure that the messages are really from different drivers on the main road, but it is not Alice's trick to deceive them away from the main road.

In this paper, we present a novel privacy-preserving vehicular announcements aggregation scheme to solve Alice's problem. It allows a leader who firstly generates a message and some responders who echo the message to issue an announcement message commonly, but the participants will be concealed in a larger group of possible users. Our scheme provides majority-based authentication and a certain level of privacy, since the receiver of the announcement is able to tell the number of participants while cannot tell the identities of them. Hence our proposed scheme satisfies the seemingly contradictory requirements in vehicular announcement scenario. Specifically, our contributions in this paper are: 1) We propose an aggregation scheme which can prevent the sybil attack beforehand, and it is different from the existing works in the mode of tracing the misbehavior afterward. To our best knowledge, it is the first scheme using threshold cryptographic technology to achieve that goal. 2) We design an interactive threshold ring signature scheme in order to meet the demands of the aggregation scheme. Different from the traditional threshold ring signature scheme in which a signer uses his private key together with the private keys of the other signers to sign a common message, it can generate a threshold ring signature without the full delegation of the private keys of the other signers or the multiple interactions between all the signers. 3) We extend our aggregation scheme to the NS-2 network simulator to confirm the availability and efficiency of our proposed scheme.

The rest of this paper is organized as follows. In Section II, we give an introduction of related works. In Section III, an introduction of the cryptographic preliminaries we used is given. The main construction of our scheme is introduced in Section IV. The security analysis and performance analysis of our scheme are given in Section V. In Section VII, the simulation results of NS-2 are provided. Then, in Section VII, we give some variants of the proposed main scheme to achieve higher security requirements and discuss some problems of the proposed aggregation scheme. Finally, we conclude this paper in Section VIII.

\section{Related Work}
Majority-based authentication is a common method to prove message reliability in VANETs \cite{CNW11}\cite{DDSV09}\cite{KWL09}\cite{VS09}. Because there are no strong trusted relationships between mobile vehicles in VANETs, in majority-based authentication schemes, only the messages confirmed by over a threshold number of vehicles are considered to be reliable. Messages aggregation in VANETs is an effective way to implement majority-based authentication and reduce the network overhead meanwhile. There are a number of works designing information aggregation protocol in VANETs. In \cite{VS09}, Viejo et al. proposed an aggregation and threshold authentication scheme using multi-signature so that the digital signatures in the network packets will be at a fixed length . Molina-Gil et al. proposed a framework of aggregation for data authentication in \cite{MCC14}. They proposed the concept of geographic zones and reactive groups to implement the detailed process of aggregation in VANETs, and used a probabilistic verification algorithm to improve the computation efficiency. In \cite{HDK13}, Heijden et al. proposed the SeDyA scheme which allows more dynamic aggregation with flexible road segmentation. Concerning the privacy issues, Zhang and Qin et al. respectively implemented privacy-preserving aggregation authentication protocol based on the pseudonym technique \cite{ZWQ11}\cite{QWD12}. In \cite{DDSV09}, Daza et al proposed a threshold-signature-based privacy-preserving announcements authentication scheme. However, in order to solve the problem of linkability in threshold signature, a group of users share the same private key. Such a design increases the risk of key compromise.

In recent years, the privacy issues in VANETs have gained widespread attentions. Owing to the notorious sybil attack, majority-base authentication is not properly functioning in privacy-preserving environment. The malicious user could pretend himself as many entities to launch the sybil attack by the abuse the pseudonyms mechanism or anonymous-group-based authentication. Some solutions with strong hardware-secure assumptions are used to mitigate the sybil attack in privacy-preserving VANETs. Kounga et al. used the secure hardware mechanism to control the generation of pseudonyms so as to prevent the sybil attack \cite{KWL09}. Qin et al. used secure RSU management to implement the pseudonyms control \cite{QWD12}. It has become an arduous task to design protocols proving reliability of messages while concealing the origin of messages with weak hardware-secure assumptions. Wu et al. used the linkable group signature to identify the sybil attack \cite{WDG10}. However, the trace operation requires two expensive pairing operations, so it is not very efficient to trace doubtable messages. Chen et al. proposed a threshold anonymous announcement scheme with direct anonymous attestation and one-time anonymous authentication in \cite{CNW11}. However, the credentials of the malicious users cannot be efficiently revoked in TAA, so the malicious vehicles attacking frequently will affect the efficiency of the scheme. In \cite{Lin13}, Lin proposed a RSU-aided scheme namely LSR mitigating the sybil attack. It supports the local sybil attack detection and can efficiently track the attacker, but it cannot function satisfactorily in areas with sparse RSUs. Besides, the utilization of trusted RSUs is not helpful to bootstrap such a system.

Different from the above works, our proposed scheme is not accordant with the idea of sybil attack detection afterwards but to prevent the adversary beforehand. The proposed scheme uses the aggregation way to issue announcement messages and the threshold ring signature preliminary to provide flexible anonymous authentication so that it can efficiently achieve the reliability, privacy and authentication simultaneously. Therefore, it is suitable for the privacy-preserving vehicular announcement scenario.

\section{Preliminaries}
\subsection{Threshold Ring Signature}
Threshold cryptographic technology was first introduced by Shamir in \cite{Sha79}, which allows a group of users to share a secret and recover the secret if over threshold users join in the recovering process. However, due to the features of the scheme violating unlinkability, it cannot be directly used in a privacy-preserving scenario. Threshold ring signature was proposed by Bresson et al. in \cite{BSS02}. A $(t,n)-$threshold ring signature can express that at least $t$ members of a group of $n$ members have signed the message while it is unknown that who the actual signers are. A threshold ring signature scheme consists of two algorithms:

\begin{itemize}
 \item T-ring-sign: On an input message $m$, a ring of $n$ members including $n$ public keys, and the private keys of $t$ members, it outputs a $(t,n)$-ring signature $\sigma$ on the message $m$. The value of $t$ as well as $n$ public keys of all ring members are included in $\sigma$.
 \item T-ring-verify: On an input message $m$ and a signature $\sigma$, it outputs either $1$ or $0$.
\end{itemize}

In traditional threshold ring signature models \cite{BSS02}\cite{RH13}, the group of actual signers in the ring is considered to be fixed and have already established a trusted relationship, so the negotiation among the actual signers and the acquisition of the private keys to execute the algorithm are not considered in the model. It just uses a series of private keys in the input of the algorithm to represent a group of determinate signers sign a message together. The traditional model does not consider how to organize the group of signers or how to put their private keys together, and it assumes that the signers have been determinate and they trust each other so the private key of any group member can be used. However, in the scenario of the vehicular announcements aggregation, the members of message issuers are nondeterministic, and the initiator of the aggregation cannot know who will join the process in advance. Besides, in the process of the aggregation neither can the announcement issuers implement several interactions between themselves, nor give the private key to the initiator. Therefore, a new protocol with the fewest times of interaction should be designed to use threshold ring signature in the vehicular announcement aggregation scenario. The new model we proposed is called interactive threshold ring signature because there is a round of communication in our model, and it consists of four algorithms:

\begin{itemize}
 \item T-ring-sign-request: On an input message $m$, a ring size $n$, and the number of actual signers $t$, it outputs a $(t,n)$-ring signature generation request $\delta$ on the message $m$. The value of $t$ and $n$ are included in $\delta$.
 \item T-ring-sign-reply: On an input message $m$, a signature generation request $\delta$ on $m$ and the private key $sk$, it outputs a signature fraction $\lambda$. The corresponding public key $pk$ is included in $\lambda$.
 \item T-ring-sign-generate: On an input message $m$, a signature generation request $\delta$ on $m$, $t-1$ signature fractions $\lambda_1, \lambda_2, \cdots, \lambda_{t-1}$ including $t-1$ public keys of the actual signers, and the private key of the signer initiating the request, it outputs a $(t,n)$-ring signature $\sigma$ including $n$ public keys.
 \item T-ring-verify: On an input message $m$ and a signature $\sigma$, it outputs either $1$ or $0$.
\end{itemize}

\subsection{Combined Public Keys}
The idea of Identity-Based Cryptography (IBC) was proposed in \cite{Sha85}, 1985. In an IBC system, a meaningful string can be used as the public key so that the user of the public key can easily learn the owner of a public key without a PKI certificate, because the public key itself may be the identity of its owner. The idea of IBC helps us to reduce the burden of certificates management, but the operations of most IBC schemes are relatively complicated. Combined Public Keys (CPK) was originally a technology of key management proposed by Nan \cite{Nan06}. In \cite{LZ08}, Liu et al. firstly used the idea of CPK to construct an efficient cryptographic preliminary implementing IBC. Zhang et al proposed an ECC-based encryption scheme with the CPK preliminary in \cite{ZLH10}. In this paper, we use this preliminary to simplify the certificate management and reduce the length of the ring signature.
\section{Proposed Aggregation Scheme}
In this section, we present our novel privacy-preserving vehicular announcements aggregation scheme, which can prevent the sybil attack beforehand by using proposed interactive threshold ring signature scheme. Our construction of threshold ring signature is derived from the Generalized Ring Signature proposed by Ren \cite{RH08}, but we implement threshold cryptography and signature negotiation in nondeterministic signers. Before delving into the detailed scheme, we first introduce the basic notion of our scheme and some basic definitions.
\subsection{Overview}
In our scheme, we focus on the vehicular announcement application. We assume that the announcement messages often have long dissemination range, but the requirement of processing time is not strict. We also assume that each vehicle is equipped with a tamper-resistant black box, which provides secure storage for secret keys and secure cryptographic computation. The black box will be preloaded with some public system parameters and private keys of the user. At last, we assume that the majority of users are honest, and the adversaries constitute a relatively small fraction of the vehicles. These assumptions are used in most of the VANET security protocols.

The basic notion of our scheme is coming from the idea of echo. Consider the following scenario: Alice wants to issue some interesting stories anonymously, and she wants everyone believes what she said is true. Therefore, Alice asks Bob if he is willing to echo what she said. As for Bob, he believes Alice's story, so he tells everyone Alice's story is true. The more people echo what Alice said, the more believable the story is. Our announcements aggregation scheme complies with the logic in which Alice issues stories, and we use cryptographic techniques to implement what Alice and Bob do. There will be one car firstly issuing an announcement message about some certain events. Then, the car asks the other witnesses help him to generate an announcement packet. If the other vehicles agree with what the leader car described and are willing to help him, they can produce a fraction of the legitimate announcement. When the leader car receives over $t-1$ (a threshold value decided by the leader car) fractions, he can finally generate an announcement packet used to warn cars far away. The identities of participants are protected by mixing with the identities of fake signers which are forged by the leader car. As for the verifiers of the announcement, they only know that the announcement is issued by over $t$ different participants, but cannot tell who they really are. The detailed scheme will be introduced in Section IV.D, and some simple notations will be explained in Section IV.C in order to simplify the description.

\subsection{Adversaries}
There are several possible attacks in a vehicular communication system, and we list some typical attacks here to help us analyze the security performance of our scheme. According to the power an adversary has, we classify the attacks in vehicular communication applications into two different types: attacks from the outside and attacks from the inside. Typical attacks from the outside include:
\begin{itemize}
  \item \textbf{Unauthorized participation} Unauthorized users directly or masquerade as a legitimate user to participate in the system, and affect the management of the system.
  \item \textbf{Message modification} The content or the source of the message is altered by malicious adversaries during the transmission.
  \item \textbf{Replay attack} Replay valid messages which were sent some time before. The adversary may avoid the authentication mechanism and bother the system, because these messages are coming from legitimate users.
  \item \textbf{Trace attack} By eavesdropping issued messages, the adversary tries to trace an entity. It infringes on the privacy of the user to bind the sensitive content of the messages and the identity of the user.
\end{itemize}

Attacks from the inside of the system mean that the adversary is more powerful. The adversary can even have a legitimate private key and certificates, but abuses the mechanism to attack the system. The upgrade attack and the sybil attack are typical attacks from the inside.
\begin{itemize}
  \item \textbf{Upgrade attack} The upgrade attack means the adversary try to convince the receiver of a message that the trust level of the message is higher than it actually is. For example, in some system, roadside infrastructures and some special vehicle (such as emergency vehicles and police car) will have higher trusted level. Therefore, the adversary may masquerade as an emergency vehicle to mislead other vehicles. In our system, the number of vehicles echoing the message is the trusted level of a certain message.
  \item \textbf{Sybil attack} The adversary tries to pretend as multiple vehicles, and may try to spread false messages to the others, and make the receivers believe that the messages come from different sources and the content of messages is true \cite{Dou02}. Because of the privacy preserving mechanisms, the receivers cannot distinguish the sources of the messages. Therefore, the adversary can mislead the other cars by making use of the contradiction between authentication and privacy.
\end{itemize}

\subsection{Roles and Type of Packets}
There are four kinds of roles in our scheme: the trusted authority denoted as $\mathcal{T}$, the Initiator of the aggregation process denoted as $\mathcal{I}$, the Replier of the aggregation process denoted as $\mathcal{R}$, and the Verifier of the announcement denoted as $\mathcal{V}$. The proposed aggregation scheme involves the following three types of packet:

\begin{itemize}
  \item \textbf{Request Packet: } When an event occurs, the Request Packet is broadcast by the vehicle who detects the event. There may be some vehicles broadcasting a Request Packet, but not all of them would receive a reply, the Reply Packet. We call the one who receives Reply Packets the Initiator of the aggregation process. The Request Packet contains event description $msg$, excepted threshold $t$, anonymous group size $r$, and related cryptographic content $\Omega$. The event description $msg$ may include event coordinates, type of the event, traffic direction, name of the road, time of the event, and so on.
  \item \textbf{Reply Packet: } Replay Packet is sent after the reception of a Request Packet by any vehicle who agrees with the content of the Request Packet and is willing to join the group to issue an announcement. We call the vehicle sending the Reply Packet the Replier. There may be multiple vehicles detecting the same event at the same time, so multiple Request Packet may be received by a Replier. If that happens, the Replier sends the Reply Packet as Algorithm 1.

\begin{algorithm}
\caption{RequestReply}
\KwIn{RequestPacketArray P, LastReply L}

Sort(P, TIME);  ~~//Sort Request Packets about the same event by time

\For{i=0;i$<$P.length;i++}
{
    \If{P[i].threshold $>$ L.threshold}
    {
        Reply(P[i]);    ~~//Send a Reply Packet

        L = P[i];  ~~//Record last replay
    }
    \Else
    {
        P[i].delete();  ~~//Ignore packet P[i]
    }
}
\end{algorithm}

  \item \textbf{Aggregation Packet: } The Initiator can produce an Aggregation Packet after receiving over $t$ Replay Packets. It provides higher evidence about the existence of the reported event. Then, the Initiator broadcasts the Aggregation Packet, and it will be forwarded by other vehicles so as to issue an announcement.
\end{itemize}

\subsection{Protocol Description}
The proposed aggregation scheme consists of seven phases, namely \texttt{Setup}, \texttt{Join}, \texttt{Event found}, \texttt{Aggregation request}, \texttt{Request reply}, \texttt{Announcement generate}, and \texttt{Announ
\\cement verify}. In \texttt{Setup} phase, the system parameters are created by the trusted authority. The keys will be preset to each legitimate vehicles firstly join the system in \texttt{Join} phase. The \texttt{Event found} phase and the \texttt{Aggregation request} phase are related to the Initiator, where \texttt{Event found} creates event announcement message, \texttt{Aggregation request} produces the Request Packet. The Repliers generates announcement fraction and send back to the Initiator in \texttt{Request reply} phase. \texttt{Announcement generate} phase generates the Aggregation Packet, and disseminates it. The Verifiers could check whether an announcement has been signed by a certain number different users in the \texttt{Announcement verify} phase. The detailed description of our scheme is given below.
\begin{enumerate}
 \item \textbf{Setup.} Let $\mathbb{G}$ be an addition group consisting of points on an elliptic curve and the order of $\mathbb{G}$ is $q$. Let $P$ be a generator of $\mathbb{G}$. Firstly, the trusted authority $\mathcal{T}$ selects $n$ secret values $x_i \in \mathbb{Z}_{q}^{*}$ randomly, and computes $Y_i = {x_i}\cdot P$ for each $x_i$, $i = 1,\cdots, n$. Secondly, $\mathcal{T}$ selects four hash function $H_0 : \{0,1\}^* \rightarrow \{0,1\}^n, n$ is the length of the output of a cryptographic hash function, for example SHA-1 $n = 160$. $H_1 : \mathbb{G} \rightarrow \mathbb{Z}_{q}, H_2, H_3: \{0,1\}^* \rightarrow \{0,1\}^l$. $\mathcal{T}$ chooses a symmetric encryption scheme over $GF(2^l)$ denoted as $E$. We define $E_k(x)$ as using secret key $k$ encrypt $x$, and $E_k^{-1}(x)$ as using secret key $k$ decrypt $x$. Define $X = (x_1,x_2,\cdots, x_n)$ as the master private key vector, and define $Y = (Y_1,Y_2,\cdots, Y_n)$ as the master public key vector. Finally, $\mathcal{T}$ sets the system public parameters to be $(\mathbb{G},q,P,Y,H,E)$ and makes them public.
 \item \textbf{Join.} The users in our system are the vehicles, and the private keys of them are issued by a trusted authority (such as transport authority) in \texttt{Join} phase. Every new vehicle will run this protocol before it joins the system. For every user with identity $ID$, the private key of the user will be computed by the trusted authority as
     \begin{displaymath}
     sk_{ID} = \sum_{i=1}^n h_ix_i \bmod q
     \end{displaymath}
     where $h_i$ is the $i$th bit of $H_0(ID)$, $i = 1,\cdots, n$. The identity of the vehicle will be permanent, for example the the license plate number. Then, the private key will be transferred to the vehicle in a secure channel. Note that each vehicle will have only one fixed private key and one permanent identity.
 \item \textbf{Event found.} Having detected an event, the Initiator $\mathcal{I}$ produces an event description $msg$. Then, $\mathcal{I}$ chooses an appropriate value of threshold $t$ and size of the anonymous group $r$ according to the current total number of vehicles in the communication range. The anonymous group is often called ring, which is used to hide the real group of signers in a bigger group.
 \item \textbf{Aggregation request.} This phase is used to produce the cryptographic content in the Request Packet by the Initiator. It includes the following steps:
        \begin{enumerate}
        \item Randomly select $r-t$ identities, and the value of $r-t$ must be higher than five. Define $\overline{S} = \{ID_1, ID_2, \cdots, ID_{r-t}\}$. For each $ID_i \in \overline{S}$, compute $PK_i = \sum_{j=1}^n {h_j}{Y_j}$, where $h_j$ is the $j$th bit of $H_0(ID_i)$, and $Y_j$ is the $j$th value in the master public key vector $Y$, $j = 1,\cdots, n$.
        \item Generate a random number $\gamma_i$ as an index for each $ID_i \in \overline{S}$.
        \item Create forgeries using each $PK_i$. Select $a_i, b_i \in \mathbb{Z}_{q}^{*}$ arbitrarily, and compute $\alpha_i = a_i P + b_i PK_i, \beta_i = -b_i^{-1} H_1(\alpha_i)$, and $m_i = a_i \beta_i$. It can be shown that $(\alpha_i, \beta_i)$ is a valid EC-Elgamal signature of $m_i$, because $m_i \cdot P = H_1(\alpha_i) \cdot PK_i + \beta_i \alpha_i$. Define $\Omega = (\{ID_1, ID_2, \cdots, ID_{r-t}\}, \\
            \{\gamma_1, \gamma_2, \cdots, \gamma_{r-t}\}, \{m_1, m_2, \cdots, m_{r-t}\}, \\
            \{\alpha_1, \alpha_2, \cdots, \alpha_{r-t}\}, \{\beta_1, \beta_2, \cdots, \beta_{r-t}\})$
        \item Wrap $\Omega$ together with the event description $msg$, threshold value $t$, and ring size $r$ as a Request Packet. Finally, $\mathcal{I}$ broadcasts the aggregation request by sending the Request Packet.
        \end{enumerate}
 \item \textbf{Request reply.} This phase is executed by the Replier $\mathcal{R}$. If the user receiving the Request Packet agrees with the event description and is willing to join the group to issue an announcement, it will run the following algorithm:
        \begin{enumerate}
        \item Parse the Request Packet as $(msg, t, r, \Omega)$.
        \item Compute a symmetric key $k = H_2(msg)$, so the size of the key $k$ is $l$.
        \item Construct a polynomial $f$ over $GF(2^l)$ such that $deg(f) = r - t, f(0) = H_3(t||r), 
        f(\gamma_i) = E_k(m_i)$, $i = 1,\cdots, r-t$.
        \item Choose random index $\gamma \notin \{\gamma_1, \gamma_2, \cdots, \gamma_{r-t}\}$, and compute $m = E_k^{-1}(f(\gamma))$.
        \item Generate a random number $c \in \mathbb{Z}_q$, and compute the EC-Elgamal signature $(\alpha, \beta)$ of $m$, where $\alpha = c P, \beta = (m - sk H_1(\alpha))c^{-1}$.
        \item Wrap $\gamma$, $m$, $(\alpha, \beta)$, and the identity of the replier $ID$ as the Reply Packet. Finally, $\mathcal{R}$ sends the Reply Packet to the Initiator $\mathcal{I}$.
        \end{enumerate}
 \item \textbf{Announcement generate.} After sending a Request Packet, the Initiator $\mathcal{I}$ waits for the Reply Packets coming from the neighbors. Once $\mathcal{I}$ receives over $t$ Reply Packets, it can generate a joint announcement. Assume the identities of the Repliers are $S = \{ID_{r-t+1}, ID_{r-t+2}, \\
     \cdots, ID_r\}$. $\mathcal{I}$ combines the signatures in the Reply Packets with the forgeries in the Request Packet to produce a threshold ring signature, which is $A = (msg,t; S \cup \overline{S}; <\gamma_1, m_1, \alpha_1, \beta_1>, <\gamma_2, m_2, \alpha_2, \beta_2>, \cdots, <\gamma_r, m_r, \alpha_r, \beta_r>)$. The Announcement Packet contains the signature will be broadcasted and forwarded to prove the existence of an event, and the real signers who participate in the process of signature generation are hidden in a bigger group to preserve their privacy. Finally, the Aggregation Packet will be broadcasted and disseminated in a wide range.
 \item \textbf{Announcement verify.} When a user receives an announcement, he can run the following algorithm to verify the truthfulness and reliability.
        \begin{enumerate}
        \item Parse the Announcement Packet as $(msg,t; <ID_1, \gamma_1, m_1, \alpha_1, \beta_1>, <ID_2, \gamma_2, m_2, \alpha_2,\\
            \beta_2>, \cdots, <ID_r, \gamma_r, m_r, \alpha_r, \beta_r>)$.
        \item Compute a symmetric key $k = H_2(msg)$.
        \item For each $ID_i, i = 1, \cdots, r$, compute $PK_i = \sum_{j=1}^n {h_j}{Y_j}$, where $h_j$ is the $j$th bit of $H_0(ID_i)$, and $Y_j$ is the $j$th value in the master public key vector $Y$, $j = 1,\cdots, n$. Verify the equation $m_i P = H_1(\alpha_i) \cdot PK_i + \beta_i \alpha_i$. If any one of the tuples $<m_i,\alpha_i,\beta_i>$ does not satisfy the equation, the verifier rejects the signature.
        \item Recover the polynomial. Randomly select $r-t$ pairs of $<\gamma_i, E_k(m_i)>$ in the received packet, and pair $<0, H_3(t||r)>$ to reconstruct the polynomial $f$ such that $deg(f) = r - t, f(0) = H_3(t||r), f(\gamma_i) = E_k(m_i)$
        \item Check if the rest of the pairs $<\gamma_j, E_k(m_j)>$ in the Announcement Packet satisfy $f(\gamma_j)=E_k(m_j)$. If any one of the pairs does not satisfy the equation, the verifier rejects the signature. Otherwise, the verifier accepts the signature.
        \end{enumerate}
        The signature convinces that there are more than $t$ participants who wants to report the event while it is not sure that who the participants really is. If the signature in the Announcement Packet is verified, the receiver can believe the truthfulness of the reported event.
\end{enumerate}
\section{Evaluation}
\subsection{Security analysis}
In this section, we informally analyze that how our proposed scheme can satisfy the requirements of unforgeability, reliability and privacy, and defend against typical attacks in the vehicular announcement scenario.
\begin{itemize}
  \item \textbf{Unforgeability:} The unforgeabilty of the proposed ring signature is based on two difficult problems. The first one is the one-way function of creating forgeries introduced in proposal \cite{RH08}. If the adversary forge a signature without breaking the verification polynomial, he is able to forge valid Elgamal signatures of any specified messages, which is considered to have negligible probability. The formal proof is referring to \cite{RH08}. The second one is the difficulty of solving a equation with high degree. It is generally known that there is no general method to solve the equation when the degree of the equation is higher than five. Therefore, it is infeasible to forge Elgamal signature pairs without a private key satisfying the specified verification polynomial. The formal proof is referring to \cite{BSS02}. Based on the two facts, The possibility of forging a signature is negligible.
  \item \textbf{Reliability:} The announcement messages are protected by digital signatures. The utilization of symmetric encryption ensures the verification polynomial used in one signature cannot be corresponded to signature of a different message. If the adversary wants to modify the content of an announcement, he has to forge a signature, and it is infeasible because of the unforeability of the signature. If the adversary wants to issue fake event report with high trust level in a legitimate way, these aggregation requests will not result in any replies, since we assume that the number of malicious vehicles is relatively small. Hence our proposed scheme satisfies the requirement of reliability.
  \item \textbf{Privacy:} We firstly consider the privacy of the multi-hop disseminated announcement packets. The Announcement Packets is the only packet which will be broadcasted and received by many vehicles in the network. The privacy of participants producing the Announcement Packets is protected by the threshold ring signature. The threshold ring signature provides indistinguishability among all ring members, but only part of them are the actual signers of the signature. It means the group of participants of the aggregation will be concealed in a larger group of possible signers. Therefore, our scheme provides a certain level of privacy. The other packets will be disseminated in one-hop communication range, but these packets leaks the privacy of the participants. The privacy preservation is embodied in three aspects. Firstly, the Request Packets and Reply Packets will be broadcasted in one-hop communication range. It means that only the participants of the event can receive the packets. Secondly, the content in Reply Packets are meaningless random strings. Hence, the identity of user will not be linked with the detailed event message. Finally, the Repliers can deny having sent the Reply Packet, because it can be a forgery made by the Initiator, so it cannot be proved that a certain one key is used in our scheme. Therefore, it is still preserving the privacy of the participants to some degree.
  \item \textbf{Prevention of sybil attack:} Our scheme applies threshold cryptographic technique based on the Lagrange Interpolation. Therefore, the signature cannot be generated unless over the threshold different private keys are used. Note that every vehicle has only one bound identity and corresponding private key in our scheme, so it is easily proved that there are more than $t$ vehicles involves in the aggregation. Meanwhile, the utilization of ring guarantees that no one is able to tell who the signers are although every user has a fixed signing key by hiding the actual signers among a group of indistinguishable fake signers. Different from the idea of existing works, the idea of threshold cryptography defends against the sybil attack beforehand, and the adversary will never launch the sybil attack without the help of the number exceeding the threshold of legitimate users.
  \item \textbf{Prevention of upgrade attack:} The upgrade attack means the adversary try to convince the receiver of a message that the trust level of the message is higher than it actually is. Specifically, in our scheme, the attack may modify the threshold value in a legitimate Announcement Packet. In our protocol, the construction of verification polynomial is related to the threshold value which the Initiator first set. Tampering the trust level of the announcement will result in verification failure, and the receiver of the announcement will drop the message. Hence the designed protocol can defend against the upgrade attack.
  \item \textbf{Prevention of reply attack:} The adversary replays the received legitimate message before in order to mislead the other vehicles about the existence of a certain event. In our scheme, the event description $msg$ generated in the \texttt{Event found} phase is required to contain the time description of the event. The replay attack can be easily prevented by checking the current time and the event time unless the adversary tampers the content of the message and forges a valid signature.
\end{itemize}
Then, we compare the security functionalities of our scheme with two relevant schemes, which are TAA \cite{CNW11} and LSR \cite{Lin13} in Table I. Data integrity and anonymity are satisfied in all proposals. Threshold verification is not considered in LSR. As group signature is used in TAA and LSR, the central group manager must be introduced and trusted. In our scheme, we use the idea of threshold ring signature, so provide flexible anonymous group assembly and threshold verification. Besides, our scheme uses message aggregation way to issue an announcement message and improve efficiency, while TAA and LSR are normal vehicular communication scheme. Finally, both TAA and LSR could implement trace the sybil attack afterward, and LSR depends on the help of RSU to implement fast misbehaviour tracing. However, our scheme can prevent the sybil attack beforehand by using threshold cryptographic techniques.
\begin{table*}
\caption{Comparison of functionalities}
\begin{center}
\scalebox{0.8}
{
\begin{tabular}{c|c c c c c c c}
  \toprule
                         &    Data   & Anonymity &  Threshold   &    Decentra-    &     Without     & Message   &    Sybil   \\
                         & Integrity &           &              &    lization     &      RSU        & Issue    &    Attack  \\
  \midrule
  TAA                    & ${\surd}$ & ${\surd}$ &  ${\surd}$   &  ${\texttimes}$ & ${\surd}$       &  normal   &   tracing  \\
  LSR                    & ${\surd}$ & ${\surd}$ &${\texttimes}$&  ${\texttimes}$ & ${\texttimes}$  &  normal   &   fast tracing  \\
  Our Scheme             & ${\surd}$ & ${\surd}$ &  ${\surd}$   &    ${\surd}$    & ${\surd}$       &aggregation&  preventing\\
  \bottomrule
\end{tabular}
}
\end{center}
\end{table*}
\subsection{Performance analysis}
In this section, we mainly analyse the performance of cryptographic operations of our scheme, and the simulation will be introduced in the next section. The proposed scheme is implemented with cryptographic library PolarSSL \cite{POLARSSL} and math library GMP \cite{GMP}. The test data were collected from a HP Compaq 8200 Elite SFF PC, which is equipped with Intel Core i5-2400 quad-core CPU, clocked at 3.10 GHz, 4 GB RAM. We use the curves NIST recommended, and they are widely used in real systems \cite{ANSI9.62}. Figure 1 gives the average computation time of three phases related to cryptographic operations in our scheme.

\begin{figure}
  \centering
  \subfigure[Ring size $r = 20$]{\includegraphics[width=0.43\textwidth]{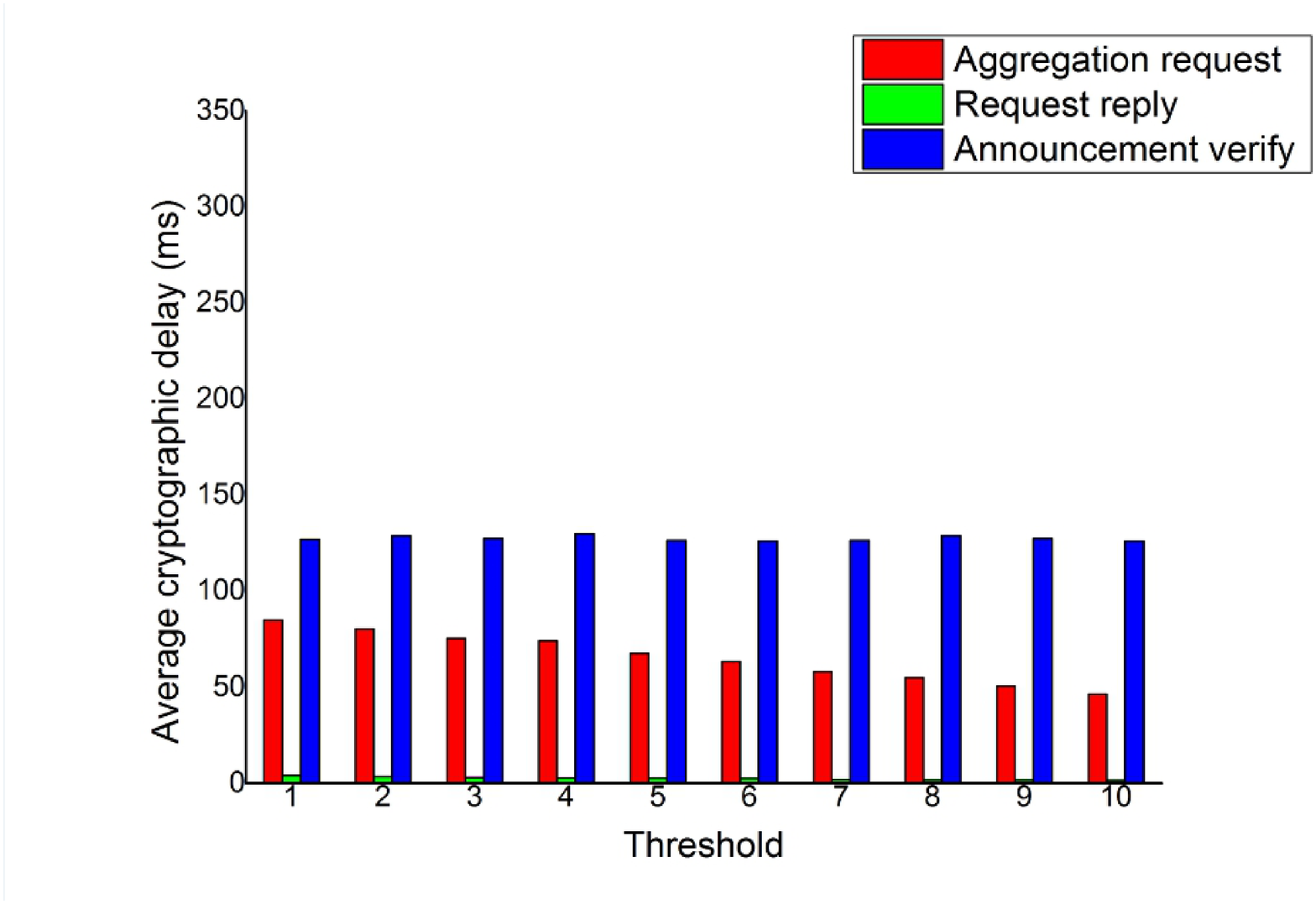}}
  \subfigure[Ring size $r = 30$]{\includegraphics[width=0.43\textwidth]{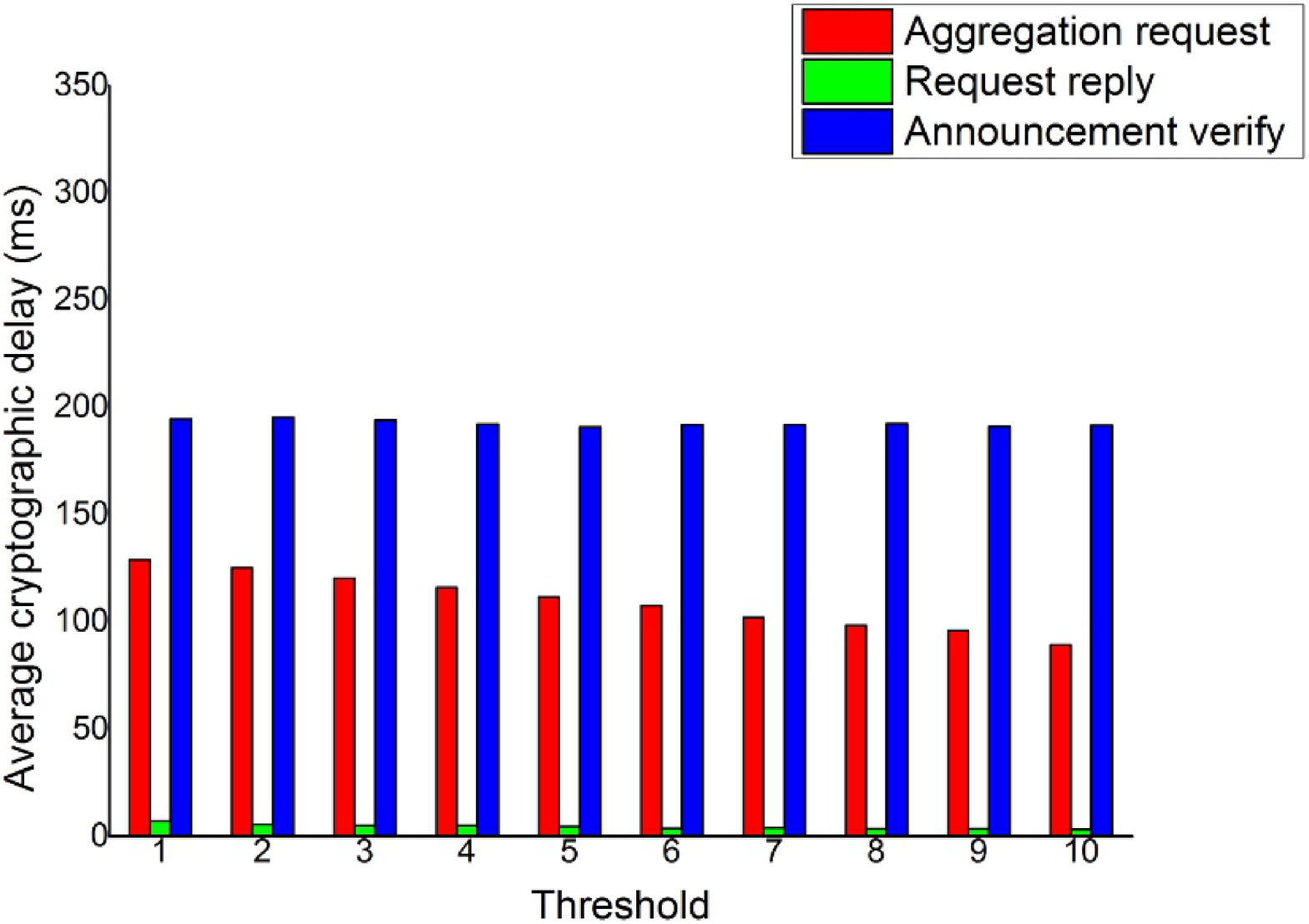}}
  \subfigure[Ring size $r = 40$]{\includegraphics[width=0.43\textwidth]{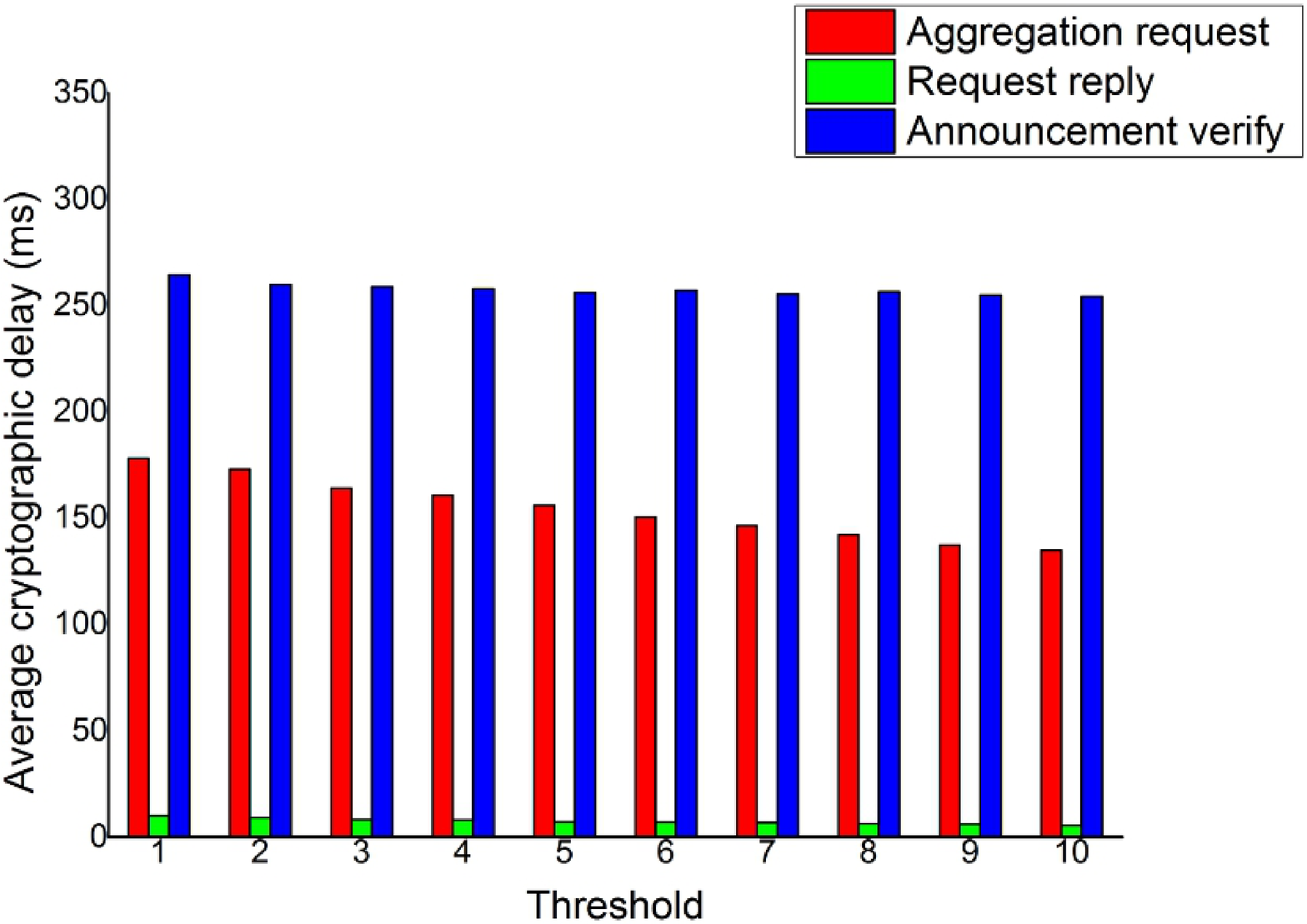}}
  \subfigure[Ring size $r = 50$]{\includegraphics[width=0.43\textwidth]{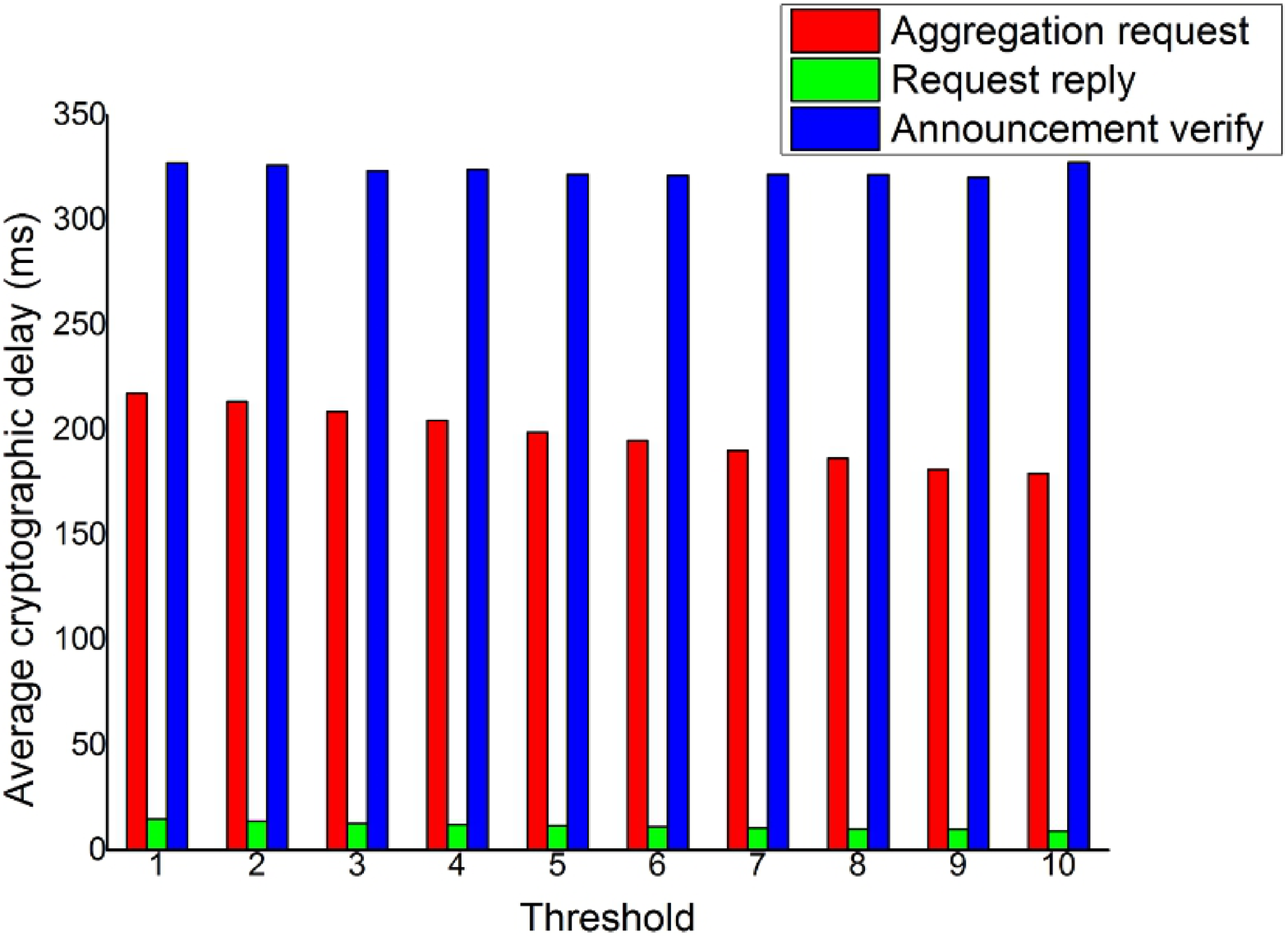}}
  \caption{Average cryptographic computation time in different phases of the scheme}
\end{figure}

In general, the most expensive operations can be computed in a few hundred milliseconds, which can satisfy the application demands in the vehicular announcement scenario. The computation time of \texttt{Aggregation request} and \texttt{Request reply} decrease with the increment of threshold value when the ring size is fixed, and the computation time of \texttt{Announcement verify} is regardless of the threshold value, and is related to the size of the ring. The reason is that the most expensive computation in our scheme is related to the number of the forged signers. When the ring size is fixed, the more actual signers participant (the threshold indicates the number of the actual signers) the less fake signers will be forged. As for the Announcement verify phase, all the signers must be treated equally because the Verifier cannot tell who the actual signers are. Therefore, it has nothing to do with the threshold value in the view of the computation overhead. The computation time of \texttt{Request reply} is very small comparing to the other two phases, so the Repliers can response the Initiator in time and efficiently finish the process of aggregation.

On the other hand, the choice of ring size and threshold is also related to the anonymity level of the ring members. There are two ways to express the anonymity in our scheme. The first one is the probability that an adversary judges if a ring member is an actual signer of the signature correctly. Obviously the success rate is the ratio of the threshold $t$ to the ring size $r$.  The other method uses the probability in which an adversary can find at least the number of actual signers of a signature. For example, consider a signature with ring size $3$ and threshold $2$, and the adversary guesses two members whom he considers are the actual signers of the signature. The probability that he guesses correctly at least one actual signer is 100\%, and the probability of at least two is $1/3$. Figure 2 shows the relationship between the number of the actual signers founded successfully by the adversary and the probability in different threshold $t$ and ring size $r$. Both of the ways show us that higher threshold setting leads to lower anonymity, although it takes less time in cryptographic computation. Repliers may not be willing to sacrifice their privacy with a lower anonymity request, so the Initiator should balance the anonymity, the computation cost and the trusted level by choosing appropriate parameters in the aggregation request.

\begin{figure}
  \centering
  \includegraphics[width=0.6\textwidth]{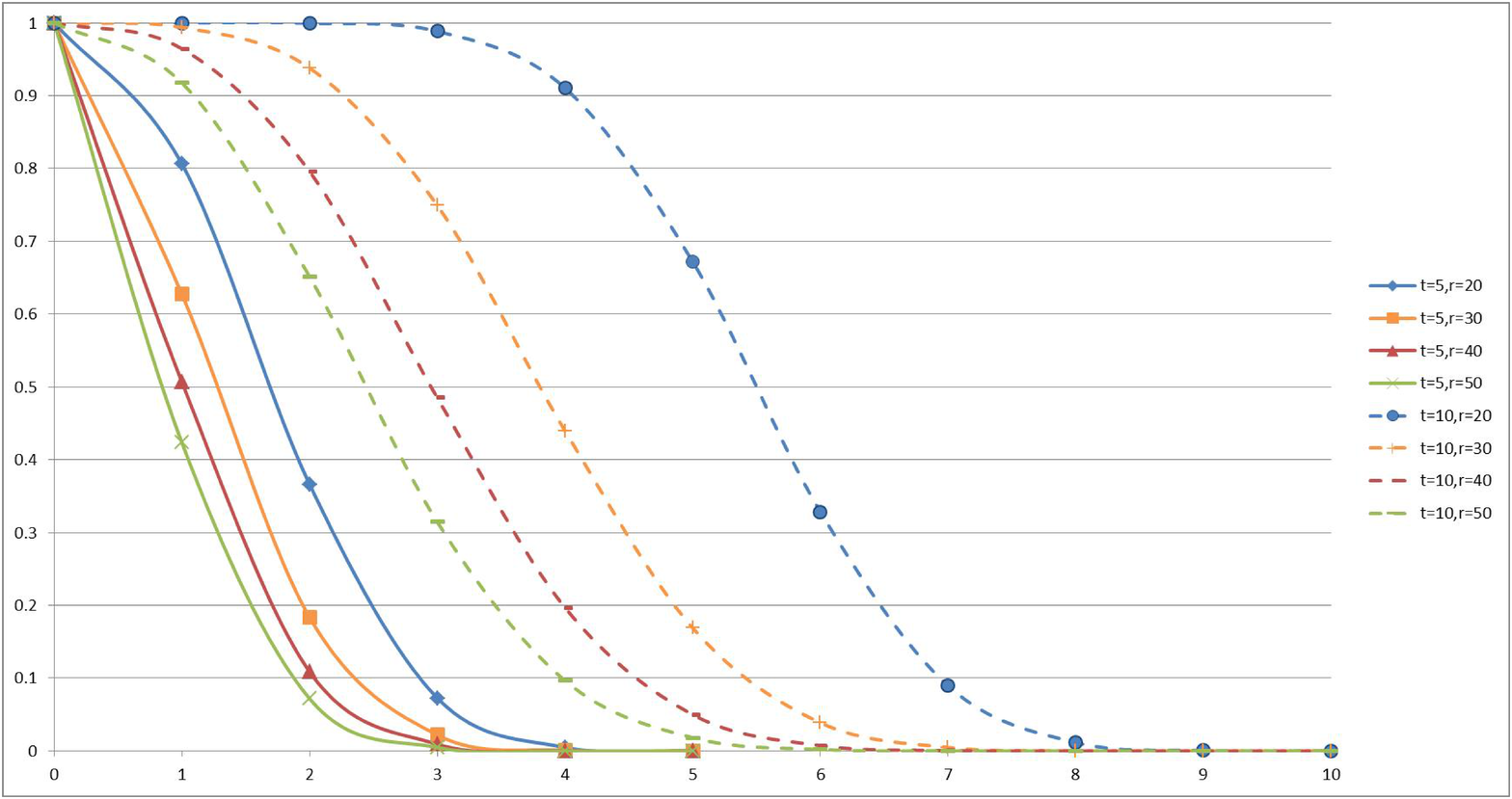}
  \caption{The relationship between the number of the actual signers founded successfully by the adversary and the probability in different threshold $t$ and ring size $r$}
\end{figure}

We then compare our scheme with the TAA scheme. The implementation of TAA is based on the PBC library \cite{PBC}.  Table II gives the comparison of cryptographic computation time at the same security level. We assume that there are 10 vehicles who want to report a same traffic jam on a certain road. In TAA, the announcements are generated independently by each vehicle, and the generation time is about 24.5ms. As for the Verifiers, they have to verify the signature of every message one by one, and it takes about 74.3ms to verify one announcement. In our example, the Verifier receives 10 announcements, so the total verification time is about 743ms. Therefore, the total cost is 767.4ms. In our scheme, the generation of an announcement is divided into two phases. The Initiator spends 46.4ms to issue a certain Request Packet in phase one, and every Replier spends 1.6ms to send a Reply Packet back to the Initiator in phase two. Because of the aggregation request and parallel replies, the generation cost of an announcement is about 48ms. The verifiers can directly verify the aggregated announcement in 126.1ms, and be sure that the message coming from at least 10 different vehicles. In our example, the advantage of announcement verification of our scheme is shown. In practice, 10 vehicles will not issue messages at the same time, so waiting enough witnesses will take more time than our analysis in TAA. However, in our scheme, the Replier who wants to echo the message will be leaded by the Initiator, so there will be less waiting time. Note that in our example the ring size is set to 20. If the Initiator wants higher anonymity, the total cost time will be more.

\begin{table}
\caption{Comparison of computation time(ms)}
\begin{center}
\begin{tabular}{c|c|c|c}
  \toprule
                     & Generation      & Verification                    & Total                        \\
  \midrule
  TAA                & $24.5$          & $74.3 \times 10$                & $767.4$                      \\
  \multirow{2}*{OUR} & Initiator: $46.4$  & \multirow{2}*{$126.1$}          & \multirow{2}*{$174.1$}       \\
                     & Replier: $1.6$  &                                 &                              \\
  \bottomrule
\end{tabular}
\end{center}
\end{table}

\section{Simulation}
In this section, we present the simulation of the proposed aggregation scheme. The simulation scenario we used was built by the tool presented in \cite{KML07}. Figure 3 shows the simulation scenario, and the others detailed parameter settings are given in Table III.

\begin{figure}
  \centering
  \includegraphics[width=0.5\textwidth]{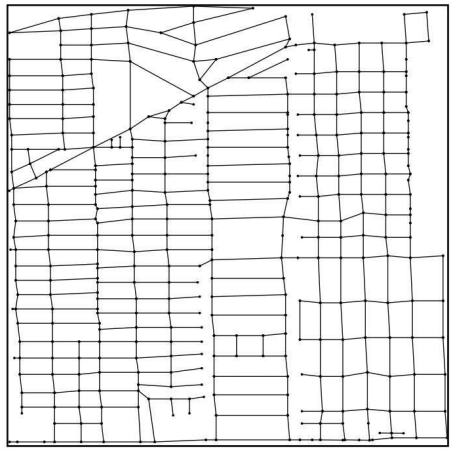}
  \caption{Simulation scenario}
\end{figure}

\begin{table}
\caption{Simulation settings}
\begin{center}
\begin{tabular}{|c|c|}
  \toprule
  Parameter & Setting \\
  \midrule
  Simulation duration & 200s \\
  Simulation area & $2400$m $\times 2400$m\\
  Number of vehicles & $50/150/250$   \\
  Average velocity of vehicles & 60km/h \\
  Vehicle Communication range & 300m \\
  MAC-layer protocol & 802.11p\\
  \bottomrule
\end{tabular}
\end{center}
\end{table}

In order to observe the degree of availability of our system, we use the NS-2 simulator to evaluate our scheme \cite{NS2}. The first indicator is the validation probability, which means the success rate in which a certain message can be endorsed by over $t$ different cars and an aggregation announcement is produced successfully. By varying the total number of vehicles in the simulation and the ring size $r$ the Initiator set, we can observe the trends of validation probability and the aggregation delay in different vehicle densities and ring sizes.

Figure 4 gives the validation probability in different vehicle densities. Vehicle density is expressed in $vehicle/km^2$. Because the ring size setting has very little effects on the validation probability (less than $1\%$ according to our simulation), we just use the condition of ring size 20 as an example shown in Figure 4. Obviously, the validation is easier for smaller threshold value or higher vehicle density.  As shown in Figure 4, the proposed scheme cannot work very well with high threshold values in very sparse networks. It is suggested to take a lower threshold value to guarantee the availability. The proposed scheme performs good in medium-density or high-density, which makes it suitable for vehicular announcement scenario. In areas with very low traffic (sparse networks), vehicular announcements are not often used, because there is hardly anybody who benefits from them.

\begin{figure}
  \centering
  \includegraphics[width=0.6\textwidth]{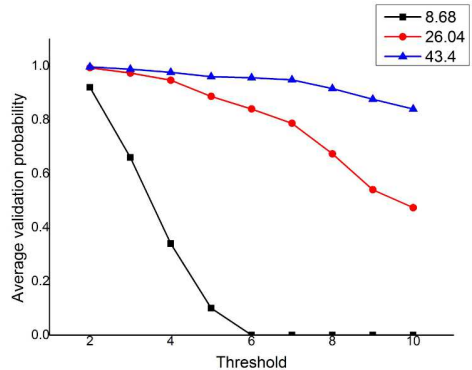}
  \caption{Average validation probability of message aggregation}
\end{figure}

There may be connectivity issues and high performance penalty during the process because there will be a round of communication between the initiator and every single responder before generating the announcement packet. Therefore, we use the interval between the time of a Request Packet sent out and the corresponding Announcement Packet produced as the second evaluation indicator so that we can learn if useful announcements can be generated in time and how the cost will be. Since the verification process mainly depends on cryptographic operations of the Verifier and with little effect of the network condition, we do not discuss the \texttt{Announcement verify} phase in this part. Figure 5 gives the simulation results of ring size 20. We only use Figure 5 as an explanation here, because the results in the other ring size settings are similar to the results of ring size 20. The line of vehicle density 8.68 rises up to $120$ms when the threshold value increases to six, because it denotes the time-out of the aggregation process. In fact, we can learn that vehicles cannot produce the Announcement Packet with low density 8.68 and high threshold value over six from Figure 4. Comparing the results in Figure 5 with the test results of cryptographic operations shown in Figure 1, we can learn that the aggregation delay is close to the computation time of cryptographic operations. Although the transmission delay and the vehicle density also affect the total consuming of producing an Announcement Packet, the magnitude of these network factors is smaller than cryptographic computations. The reason is that the process of response is very simple and fast, and the influence of the dynamics of VANETs and transmission delay is very small on the phase. In summary, the biggest impact factor of the total aggregation consuming is the cryptographic computation delay.

\begin{figure}
  \centering
  \includegraphics[width=0.6\textwidth]{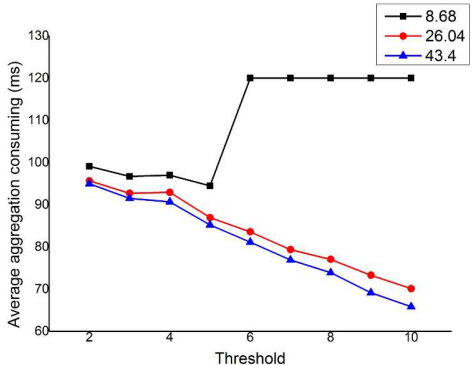}
  \caption{Average time of producing an Announcement Packet}
\end{figure}

We then further study the relationship between non-cryptographic delay and the arguments of our scheme. Figure 6 and Figure 7 depict the non-cryptographic delay, which mainly includes transmission delay, and the delay of waiting replies. The vehicle density mainly influences the delay of waiting replies. Figure 6 shows that the non-cryptographic delay is smaller when the vehicle density is higher. The reason is that the vehicle is more likely to find companions in higher density scenarios, so the aggregation is likely to be done in a shorter time. The ring size mainly influences the transmission delay in communications, because the network packet size is related to the ring size setting. Figure 7 shows that the non-cryptographic delay will increase according to the increment of the ring size of the announcement. The threshold value also contributes to the non-cryptographic delay. We can learn that the delay increase linearly with the increment of threshold value from both Figure 6 and Figure 7. The reason is that the waiting time of every Repliers can be seemed as obeying the uniform distribution, so the total waiting time emerges the linear relationship to the number of participants.

\begin{figure}
  \centering
  \includegraphics[width=0.6\textwidth]{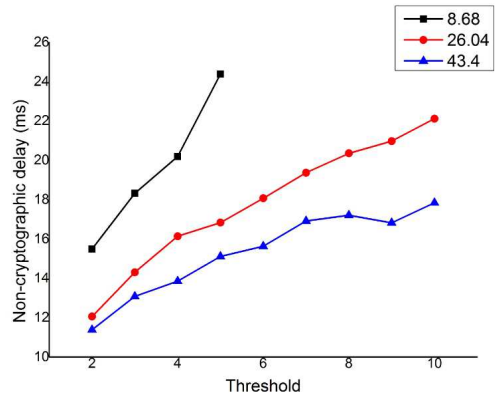}
  \caption{Non-cryptographic delay in different vehicle densities}
\end{figure}

\begin{figure}
  \centering
  \includegraphics[width=0.6\textwidth]{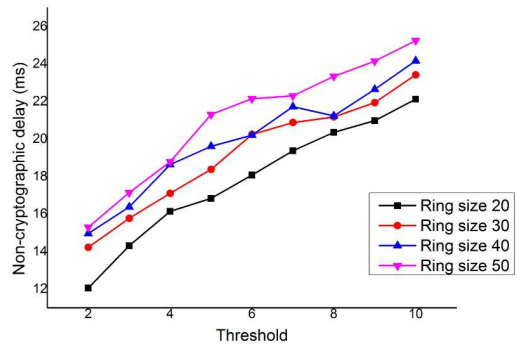}
  \caption{Non-cryptographic delay in different ring size settings}
\end{figure}

\section{Variant and Discussion}
In this section, we introduce two variants of the proposed scheme to achieve higher security goals and discuss some possible problems of our scheme.

As discussed in \cite{HLH12}, the utilization of CPK may involve a kind of collusion attack. The basic idea of the collusion attack is that some adversaries who have legitimate private keys commonly compute the secret master private key vector. If enough ID linearly independent adversaries use their private key values to carry out the collusion attack, they will get part of or maybe all of the content of the master private key vector. Then, they can calculate the private keys of other innocent users. In our basic scheme, we did not consider this kind of attack, because the utilization of vehicular black box can protect the private key from being disclosed. It means even the adversary has a legitimate private key, he cannot abuse it or get its value. However, in order to avoid the hardware-based assumption, we give a variant of our main scheme to achieve higher security level in this section.

In \texttt{Setup} phase, the trusted authority $\mathcal{T}$ computes the private key of user with identity $ID$ as $sk_{ID} = \sum_{i=1}^n h_ix_i + \mu \bmod q$, where $\mu$ is chosen uniformly from $\mathbb{Z}_{q}^{*}$. Both the private key $sk_{ID}$ and random value $\mu P$ are assigned to the user. In \texttt{Aggregation request} phase, the Initiator $\mathcal{I}$ computes each public keys of fake ring signers as $PK_i = \sum_{j=1}^n {h_j}{Y_j} + \nu P$, where $\nu$ is a random element in $\mathbb{Z}_{q}^{*}$ chosen by $\mathcal{I}$. The forgeries created by $\mathcal{I}$ are changed to $(\alpha_i, \beta_i, D = \nu P)$. Also, the signature computed by the Replier $\mathcal{R}$ in \texttt{Request reply} phase is changed to $(\alpha_i, \beta_i, D = \mu P)$. In \texttt{Announcement verify} phase, the Verifier $\mathcal{V}$ check the equation $m_i P = H_1(\alpha_i) \cdot (PK_i+D) + \beta_i \alpha_i$ for each $ID$ in the ring signature.

In our proposed variant, the introduction of random value in private key assignment can mitigate the collusion attack by increasing the unknown numbers. However, the length of the final announcement is increased to approximately 4/3 times of the original version. Although the simulation results show that the size of the network packets will not significantly affect the efficiency, another simple method of increasing the size of the key vectors also helps to mitigate the collusion attack to some degree. We present the variant to the discussion part, because the significance of the variant version may be depending on the application environment.

Another problem is the privacy in one-hop communication range. In our basic scheme, the Repliers will leak its identities to its neighbors in one-hop communication range. The privacy preservation between event witnesses is not considered, but the negotiation result (the announcement) is anonymous. We consider it is acceptable for normal-power adversaries and not very strict anonymity level. In fact, some applications (e.g safety applications) in VANETs themselves require information about one-hop neighbors \cite{AS13}. As for the normal-power adversaries, they cannot acquire identity-related information unless the adversary is one of the litigants. More powerful adversaries may be capable of eavesdropping whole networks. We present another variant with normal public key encryption mechanism to prevent eavesdropping adversaries here.

In \texttt{Aggregation request} phase, the Initiator $I$ generates a short-term key pair denoted as $pk, sk$. They can be key pairs in any kind of public key system, but we recommend ECC-based encryption for short length of ciphertext. The Initiator puts the public key $pk$ into the Request Packet, and broadcasts the packet. In \texttt{Request reply} phase, the Repliers $R$ constructs the polynomial using $f(0) = H_3(t||r||pk)$ and encrypts their Reply Packets with the public key $pk$ in the Request Packet. Finally, only the Initiator $I$ owning private key $sk$ can decrypt the Reply Packets and produce the Announcement Packet.

The eavesdropping adversaries cannot read the Reply Packet because of encryption, so they will not learn the identity of the participants by monitoring the negotiation process. If the adversary changes the public key value $pk$ in \texttt{Aggregation request} phase, it will be equivalent to re-initiate a new aggregation request of the adversary. The reason is that the broadcasted Request Packet cannot be deleted by the adversary, and the Reply Packets are bound to the specified aggregation request. Nevertheless, the variant version of our scheme is still inadequate to guarantee the privacy in one-hop communication range when the Initiator itself is a malicious adversary. In usual cases, the Initiator issuing accurate messages is trustworthy by other vehicles. Besides, the adversary will just affect a small part of the network in this situation. Therefore, we just discuss it in this section. We think it is a difficult problem to establish trusted relationships efficiently among totally intrusted nodes in the unstable aggregation scenario. It is beyond the scope of this paper, and we will leave it as a problem for further researches. Possible solutions to account the malicious Initiator may be rejecting to reply the aggregation when the information is really sensitive to the Replier or only trusted vehicles can be elected to be an Initiator.

\section{Conclusion}
In this paper, we have presented a novel privacy-preserving vehicular announcements aggregation scheme in VANETs. The proposed scheme is based on the interactive threshold ring signature which can commonly generate a threshold ring signature between nondeterministic actual signers without the full delegation of the private keys of the other signers. Our scheme is capable of preventing sybil attack beforehand, but not providing the detection afterwards like existing schemes. Through extensive evaluation, we have demonstrated that the proposed scheme can implement the high trustworthiness and the privacy of the vehicular announcement simultaneously, and can efficiently work in the vehicular announcement scenario.

\end{document}